# Beyond-Tomorrow—Electrification of Hard-to-Abate Sectors

Jonas Huber[1], *Senior Member, IEEE,* and Johann W. Kolar[2], *Fellow, IEEE*

[1] Power Electronics and Drive Systems Laboratory, ETH Zurich, Switzerland
[2] Advanced Mechatronic Systems Group, ETH Zurich, Switzerland

***Abstract*—Reaching net-zero greenhouse-gas (GHG) emissions requires deep electrification beyond ground transportation, buildings, and conventional industry automation. This article discusses direct electrification pathways for hard-to-abate sectors, which today account for 30% of global GHG emissions and include steel, cement, chemicals, aviation, and shipping. Whereas indirect electrification via hydrogen and synthetic fuels will remain indispensable for some applications, direct use of electricity is generally preferable from an efficiency perspective. We review selected emerging opportunities for power electronics in industrial process electrification, including steelmaking through hydrogen plasma smelting reduction or molten-oxide electrolysis, electrified cement production, high-temperature process heat, plasma-assisted chemistry, and electrochemical synthesis. Across these examples, power converters must interface with highly application-specific loads, ranging from low-voltage high-current electrolysis cells to dynamic plasma reactors and megawatt-scale induction heating systems. In addition, high-power grid interfaces and in-plant energy management/buffering solutions must be clarified. Thus, future electrified industrial processes open new "beyond-tomorrow" research challenges for power electronics in close interdisciplinary collaboration with scientists from neighboring disciplines.**

## I. Introduction

The world's current energy mix is still largely dominated (80%) by fossil fuels [1], which have fueled the rapid economic growth and infrastructure build-up since the industrial revolution. Burning fossil fuels, however, comes with immediate negative consequences such as air pollution and severe long-term risks due to climate change. Limiting the average global temperature increase by the end of the century to below +1.5 °C…+2.0 °C with respect to pre-industrial temperature levels requires a rapid reduction of global greenhouse gas emissions to net-zero by 2050, i.e., within the next 25 years [2].

This implies the need for a rapid transition to an ideally fully renewable energy supply and, possibly, nuclear energy using next-generation reactors. These energy sources typically provide electricity, and there are already existing and clear pathways towards full electrification for many important end-use energy services such as road mobility (electric vehicles), heating of buildings (heat pumps), and industry automation (variable speed drives). In all cases, power electronics is a key enabling technology, with the share of electric energy processed by power converters increasing towards 100% for a fully renewably supplied energy system. Accordingly, experience curve effects continue to drive massive improvements in performance indicators like efficiency, power density, and overall performance-to-cost ratio; on the other hand, lifecycle environmental impacts and end-of-life scenarios should be considered more closely [3].

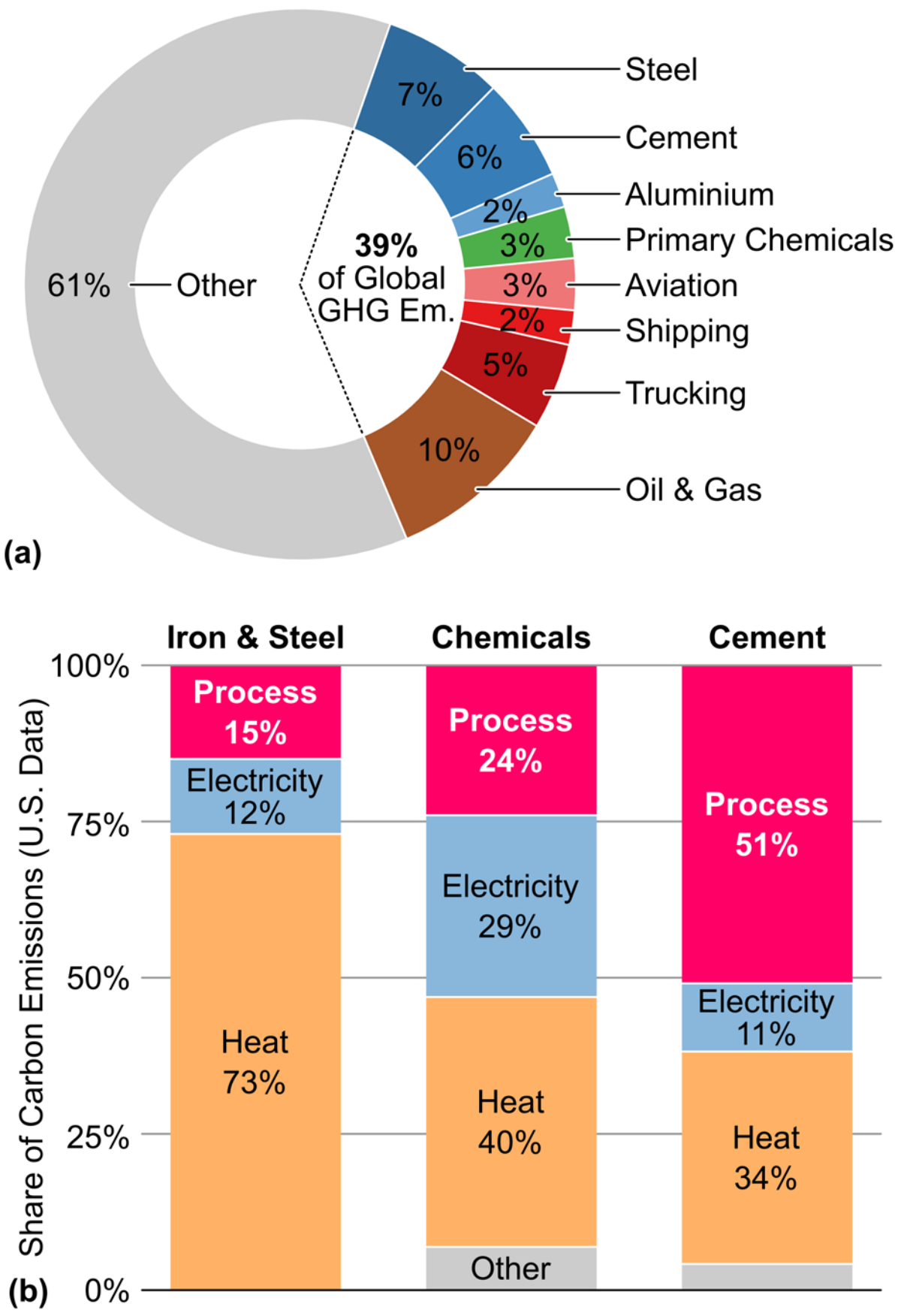

**Fig. 1. (a)** The so-called hard-to-abate sectors account for around 30% of today's greenhouse gas (GHG) emissions (data from [4]). **(b)** Carbon emissions of key sectors are dominated by burning fossil fuels for high-temperature heat, and by process emissions that result from the chemical reactions employed today (figure redrawn from [5], based on U.S. DOE data).

However, today, the so-called hard-to-abate sectors contribute about 30% of the global greenhouse gas (GHG) emissions, see **Fig. 1a**. These sectors rely on fossil fuels either because of their need for high energy density (aviation, shipping), or the need for high-temperature heat with temperatures far beyond the 200 °C that can be achieved with heat pumps. Some large-scale industrial processes (e.g., making of steel and cement) involving materials cannot be decarbonized but de-fossilized only, i.e., the $CO_2$ emissions are partly inherent to the chemical processes and cannot easily be avoided, see **Fig. 1b**. Therefore, a future energy system will not be all-electric but involve "electrons and molecules" [6]. Fed almost exclusively by (renewable) electricity, such a net-zero emissions multi-carrier energy system [7], [8] will rely on carbon capture and storage (CCS) and power-to-X technologies for providing synthetic fuels based on hydrogen (e.g., ammonia, methanol) for industrial processes that, today, do not lend themselves to direct electrification, and long-haul transportation (aviation, shipping), as well as long-term/seasonal energy storage.

However, from an efficiency perspective, direct use of electric energy in end-use energy services is clearly preferable to indirect electrification via hydrogen or synthetic fuels [9]. Therefore, this article gives an overview on pathways for the *direct* electrification of hard-to-abate sectors, with a particular focus on industrial processes. Building on an earlier white paper [3] and a keynote presentation at IEEE COMPEL 2026 [10], we highlight opportunities and future research directions for power electronics in enabling and accelerating this transition.

## II. Direct Electrification of Hard-to-Abate Industry Sectors

What follows is an overview of selected salient examples of electrification technologies, both established and emerging. The aim is not to provide a comprehensive review, which would be impossible in a single article; instead, the aim is to provide entry points for power electronics engineers to dig deeper into interesting topics to identify new research challenges. First, we discuss key applications by sectors (transportation, steel, cement) followed by a more general description of high-temperature process heat generation, before addressing electrified chemistry and, finally, system-level challenges.

### A. Transportation

Whereas battery-electric freight trucks are emerging, enabled by megawatt charging [11], there are also successful examples of direct electrification of maritime transport: In Scandinavia, electric ferries operate on up to one-hour itineraries using around 4.5 MWh batteries [12]. Even without considering economic viability, the weight and space requirements of the batteries limit this approach to short, coastal routes and is not applicable to long-haul freight shipping that moves more than 80% of all goods traded worldwide [13]. Similarly, there is a wide range of initiatives regarding electrified air transport, but except for small short-range aircraft, these concepts typically rely on (liquified) hydrogen for long-range operation [14].

Therefore, production of hydrogen using renewable electricity through electrolysis remains relevant, especially as a first step towards synthetic fuels like ammonia that can be stored and transported more easily [15]. Research in chemistry investigates methods for enhancing well-known water electrolysis, e.g., by using pulsed dc currents to improve the process efficiency [16], or consider plasma-assisted electrolysis [17]. Such emerging reactors require suitable power supply units, opening interesting R&D vectors in power electronics.

### B. Steel

The steel industry produces almost 2 billion tons of steel per year, and today's production processes release almost 2000 kg of $CO_2$ per ton of steel [18]. Conventionally, a two-step process transforms iron ore to steel: First, blast furnaces reduce iron ore to so-called pig iron, still containing a relatively high carbon content of around 4% to 5%, whereby burning coke provides the necessary carbon as well as heat (see **Fig. 2a**). This process inherently releases $CO_2$, the fuel burned for heating the blast furnace contributes to the emissions, and the preparation of the raw materials (iron ore pellets/sinters, coke) are energy-intense, too. In a second step, an oxygen converter removes excess carbon from the molten pig iron by injecting pure oxygen until the desired carbon concentration is reached (less than 2% for steel by definition). Again, this process releases CO and $CO_2$ as waste gases. Interested readers can find an in-depth process description in [19].

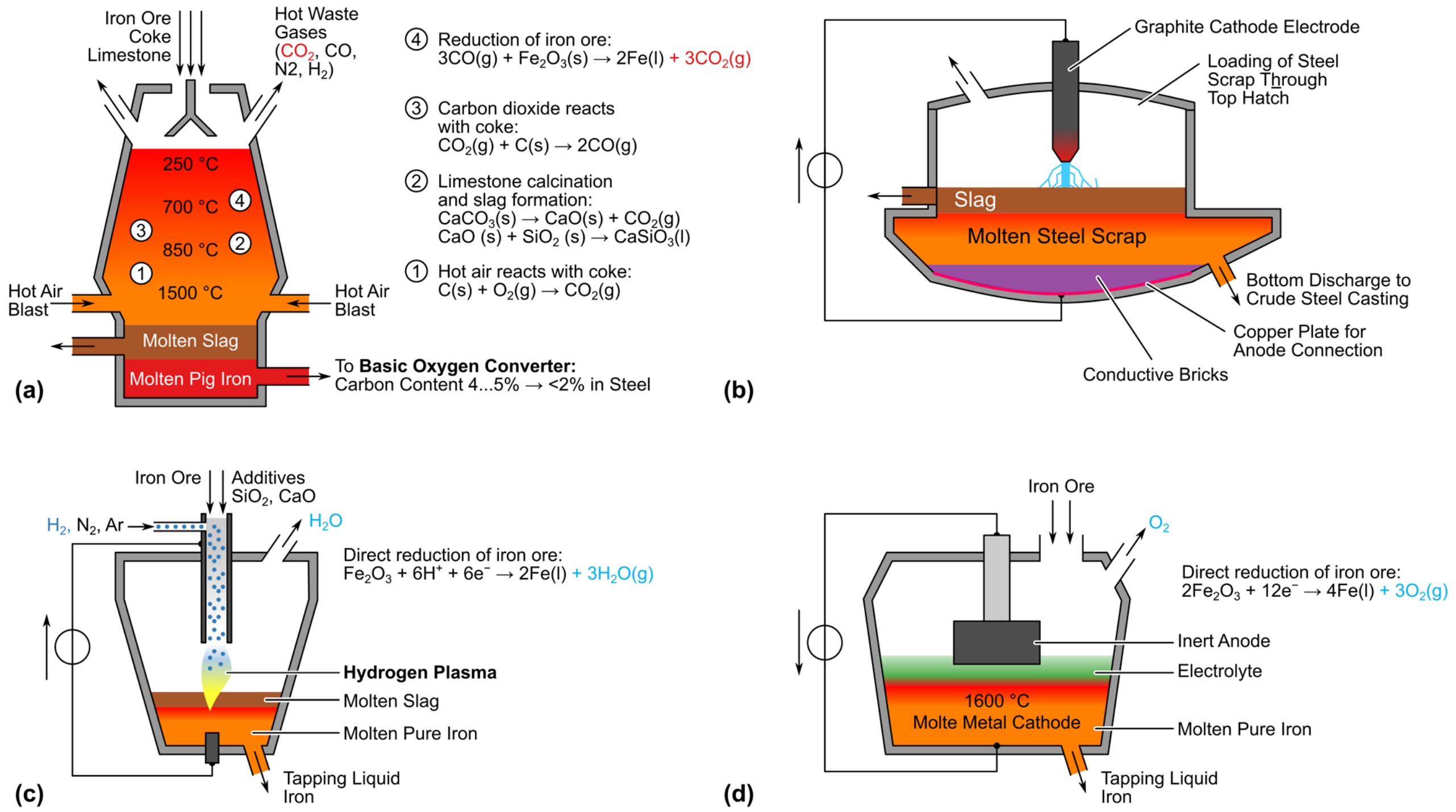

**Fig. 2.** **(a)** Blast furnace for reduction of iron ore into pig iron with relatively high carbon content (figure based on [20]); note that the involved chemical reactions inherently release $CO_2$, which comes on top of the heating-related GHG emissions. **(b)** DC Electric arc furnace (EAF) for melting of steel scrap (figure based on [21] and [19]). **(c)** Direct hydrogen plasma smelting reduction (HPSR) of iron ore with a dc transferred thermal hydrogen plasma, resulting in water vapor as the byproduct (figure based on [22], [23]). **(d)** Direct reduction of iron ore using molten-oxide electrolysis (MOE) with an inert anode, where the only byproduct is high-purity oxygen (figure based on [24]).

**Electric Arc Furnaces.** Today, around 30% of the global steel supply is obtained from recycling steel scrap, and with increasing scrap availability, this share is expected to grow to above 60% by 2050, whereby active research is still addressing the handling of contaminants in the scrap for producing high-performance steel [18]. Steel recycling uses electric arc furnaces (EAFs) to melt steel scrap. Modern EAFs use dc arcs because of better energy efficiency, lower electrode consumption, and better grid friendliness (higher power factor, less flicker) compared to legacy three-phase ac arc designs [21], [25], [26]. In such dc EAFs (see **Fig. 2b**), a transferred arc burns between a water-cooled hot cathode electrode and a molten metal bath anode [26]. For power levels above around 10 MW, multiple individual cathodes are used. Alternatively, consumable (1...1.5 kg/t steel) graphite electrodes support currents of 100 kA and power levels beyond 100 MW [26]. The arc voltage is typically relatively low (below 1 kV even for several 10 kA [26]) and depends on the length and temperature of the arc, which is moving with very high dynamics, and the voltage of the current flowing through the slag layer (depending on the slag thickness, composition, and temperature). Both voltage contributions show non-linear dependencies on the arc current, and all parameter variations must be considered in the design of the EAF power supply [21]. The power supply further requires output dc inductors to stabilize the arc, and protection systems against flashovers to the furnace walls [21]. **Fig. 3** shows exemplary high-power EFA power supply systems.

**Direct Hydrogen Plasma Smelting Reduction.** Coming back to primary iron production from iron ore, hydrogen-based direct reduced iron (H2-DRI) production uses hydrogen as the most promising alternative reducing agent, which avoids the introduction of carbon into the resulting iron, the use of coke, and hence the process-related emission of $CO_2$ [29]. Taking this one step further, employing hydrogen *plasma* improves the reaction and directly provides the necessary thermal energy (no external heating is required), reducing the heat losses and hence lowers the energy intensity of the process [29]. Advantageously, the only byproduct of the core reaction is water vapor; no carbon is released. **Fig. 2c** shows an exemplary hydrogen plasma smelting reduction (HPSR) reactor [22], [23], using a dc transferred thermal plasma between a hollow graphite cathode and the molten iron oxide. The process gas not only contains hydrogen, but also argon and nitrogen for optimizing the arc properties. Interested readers are referred to [29] for further examples and an in-depth discussion of hydrogen plasma applications in steelmaking and to [30] for a broader coverage of hydrogen plasma applications in general metallurgy. Due to the similarity between plasma torches and EAFs discussed above, similar requirements for the power supplies result; **Fig. 4a** shows a typical power supply for a high-power plasma torch. Finally, note that research also investigates iron ore reduction using non-thermal hydrogen plasma generated with microwaves, but scaling-up to higher power is challenging [29].

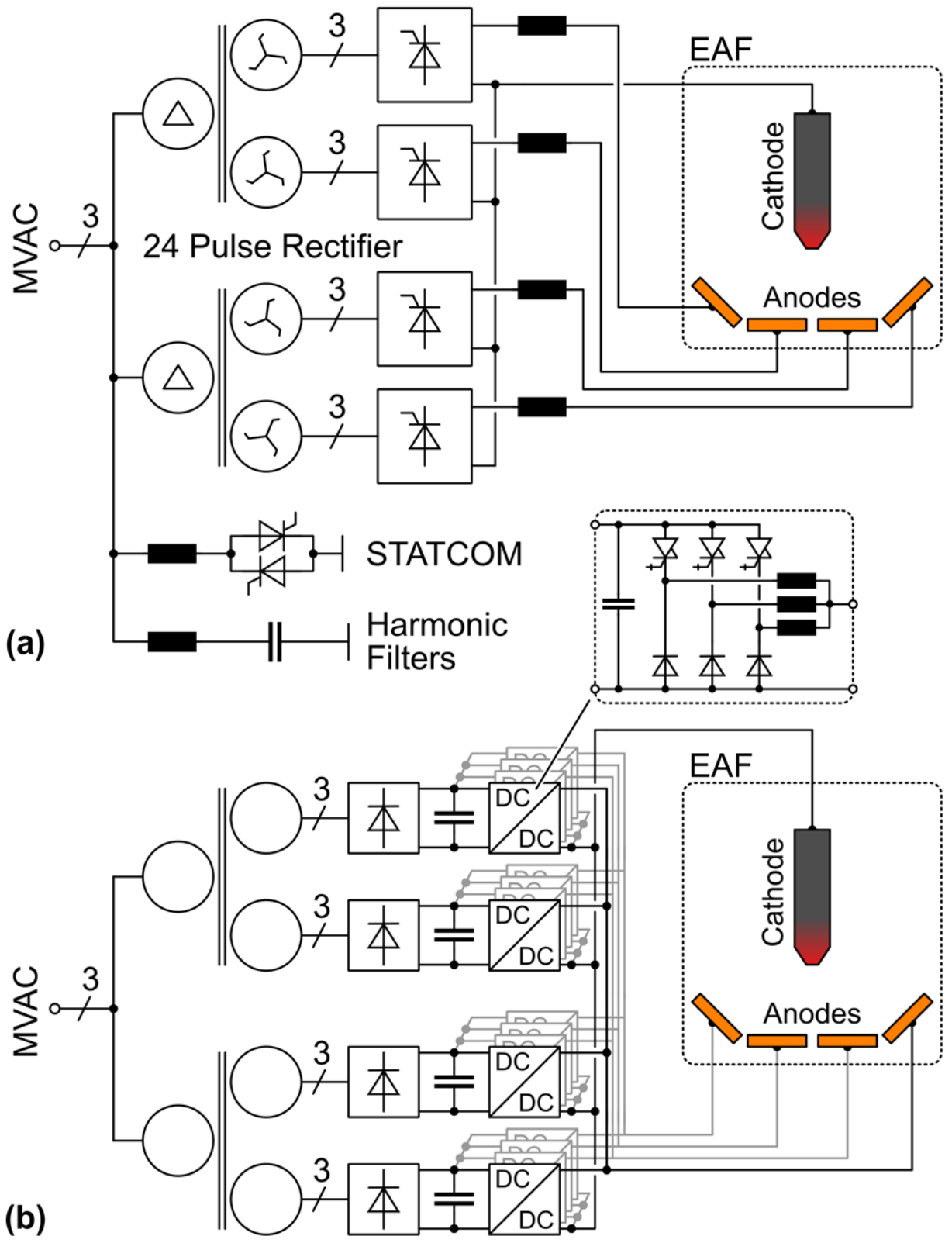

**Fig. 3.** Exemplary high-power (100 MW class) EAF power supplies. **(a)** Typical solution based on 24-pulse thyristor rectifiers and constant arc current control: The fluctuating arc voltage leads to active and reactive power fluctuations, which must be compensated by static VAr compensation and static synchronous compensators (STATCOM) to prevent flicker. **(b)** Alternative approach employing IGCT-based dc-dc choppers, enabling constant power control of the arc. Note that this topology proposed in 2005 is essentially similar to the "Panama" rectifier widely used for MVac-LVdc datacenter power supplies in China today [27]. Both figures are based on [28], where more interesting details are available.

**Molten-Oxide Electrolysis.** Finally, direct electrification of iron ore reduction is also possible via electrochemical processes, where an externally supplied electric current provides the electrons required to reduce the iron oxide, see **Fig. 2d**. Only pure oxygen results as a byproduct and no carbon is released. Specifically, molten oxide electrolysis (MOE) using special electrolyte compositions [31], and, in particular, an inert electrode material for the anode [32], has already made the leap from the laboratory to scaling up towards industrial use [24]. Requirements for the power supply are similar to other electrolysis processes (e.g., hydrogen production, aluminum smelting), i.e., high electrolysis cell currents in the kiloampere range and low cell voltages of around 2.6 V, and here, specifically, high MOE cell operating temperatures of around 1600 °C [33]. Alternative electrochemical pathways target much lower cell temperatures (100 °C) using alkaline conditions, but are still in the research state [34].

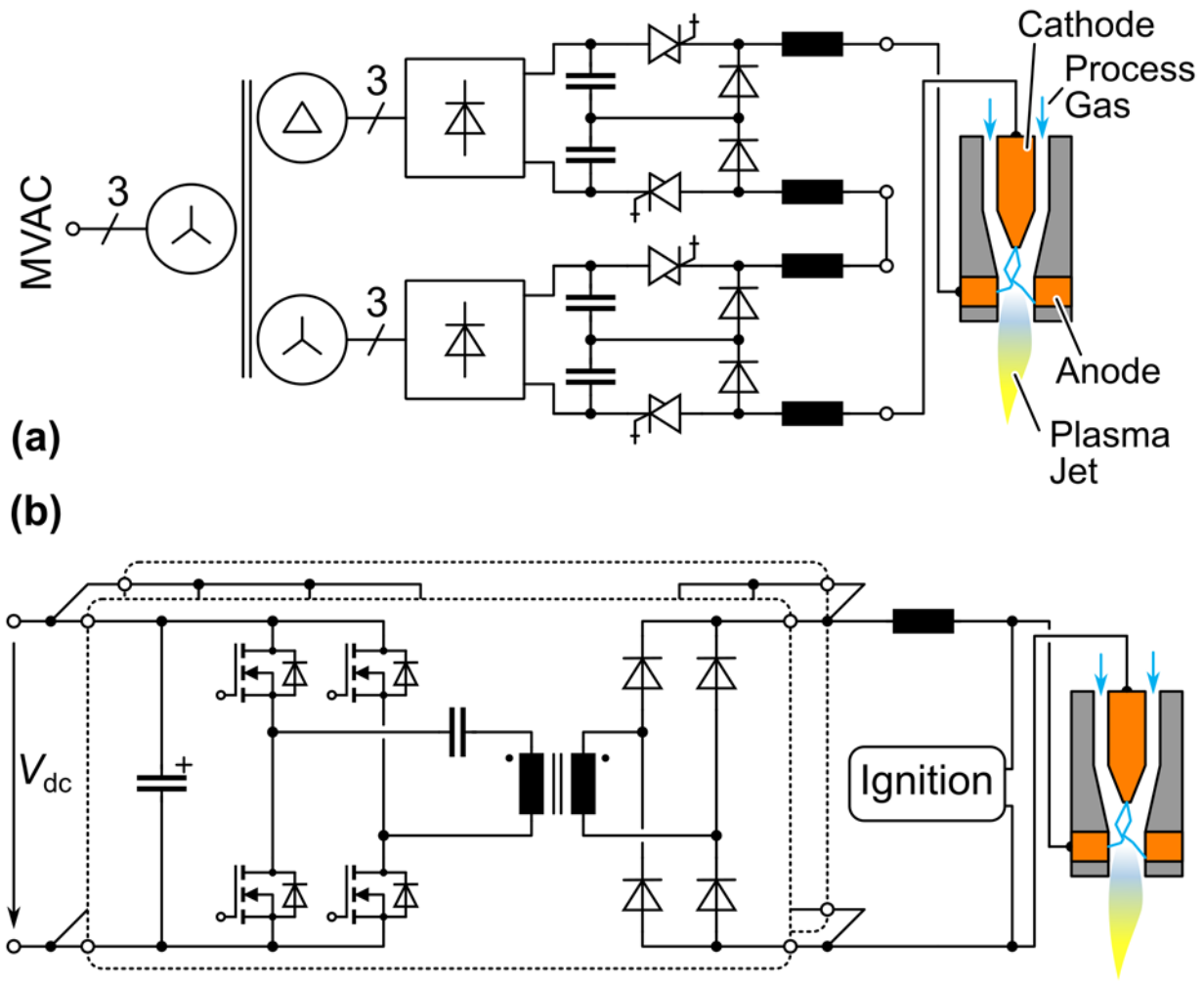

**Fig. 4.** Exemplary plasma torch power supplies for **(a)** high-power (3 MW, 2 kA) plasma torches [35] (2008), using a 12-pulse rectifier and two three-level IGCT-based choppers with series-connected outputs (note the similarity to the EAF power supply in **Fig. 3b**), and **(b)** low-power (20 kW, 80 A) torches [36] (2026), using interleaved LLC dc-dc converters realized with 1200 V SiC technology and a dedicated ignition module that provides high-voltage pulses for arc ignition; note that the high-frequency isolation approach is conceptually similar to [37] (1997).

## C. Cement

Around 4 billion tons of cement are produced annually to satisfy the global demand for around 30 billion tons of concrete [38], [39]. **Fig. 5a** visualizes a conventional plant (for further details refer to [40], [41]) producing cement clinker: First, the raw materials (mainly ground limestone and clay) are heated in a multi-stage preheating tower before entering the calciner. There, the mixture is heated further to around 900 °C, triggering the calcination reaction that breaks down limestone into quicklime (CaO) and carbon dioxide. Finally, quicklime and additives like silica react in a rotary kiln and the mixture is heated (again by burning fossil fuels, typically pulverized coal in burners with thermal output power levels of up to 250 MW [42]) to around 1400 °C to sinter into clinker. This conventional production method releases about 600 kg of $CO_2$ per ton of cement clinker [39], of which about 40% are from heating by burning fossil fuels, but the other 60% are from the process itself (calcination of limestone) and cannot be avoided [41].

However, as an alternative to using carbon-neutral (synthetic) fuels, direct electrification of the heat generation is an interesting pathway. As shown in **Fig. 5b**, heating of the calciner can be provided by resistive heating or induction heating [40]; even electric arcs are considered in a pilot factory in Sweden that is scheduled to open in 2027 [43]. Advantageously, electric heating of the calciner results in a high-purity $CO_2$ waste gas stream that facilitates effective carbon capture and use, e.g., in carbon-negative construction materials [43] or for further processing into multi-carbon feedstock (see **Fig. 9** and discussion there). Finally, plasma torches are used to achieve the high temperatures required in the rotary kiln [44], [45], with a 300 kW plasma-heated kiln using $CO_2$ as plasma process gas (in a closed cycle) demonstrated in 2025 [44]. Microwave heating has also been considered but not yet emerged beyond the research state [45].

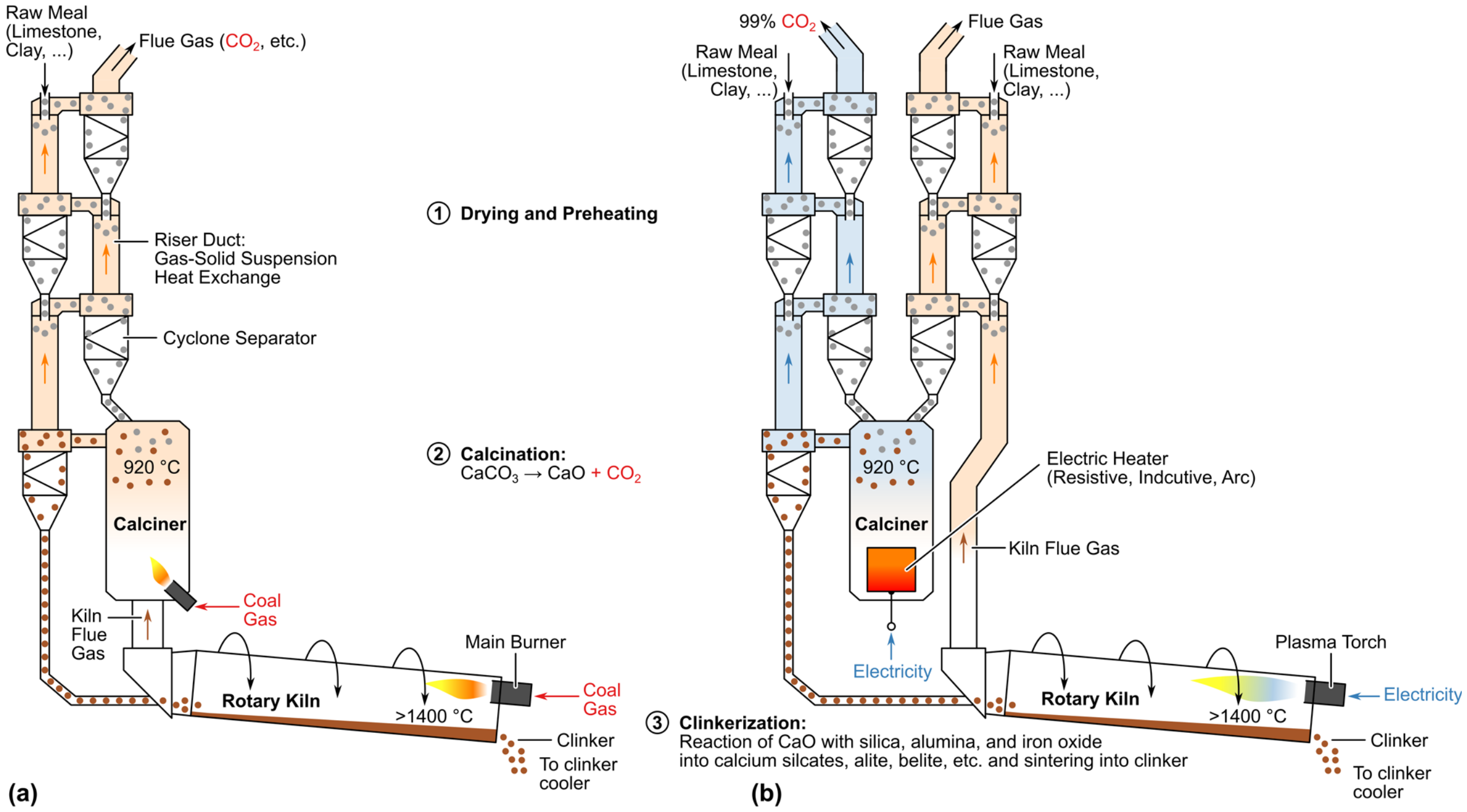

**Fig. 5. (a)** Conventional fossil-fuel-fired (typically coal) cement plant and **(b)** electrified alternative using an electrically heated calciner and a plasma torch for heating the rotary kiln (figures based on [40]).

## D. High-Temperature Heat

Process heat has featured prominently in the two large-scale process examples discussed above, and high-temperature heat demand appears in many more industrial applications, see **Fig. 6a**. Conventionally, such high temperatures are reached by burning fossil fuels. The following summary of electric heating technologies is based on [47], [48] and necessarily kept short; interested readers are referred to [47] for a technology status overview and to [48] for in-depth technical analyses of the key technologies.

**Heat Pumps.** Whereas heat pumps provide a highly efficient electrification pathway for low-temperature heating, e.g., for buildings, the maximum achievable temperatures are limited to 200 °C, see **Fig. 6b**. Power ratings in the 10s of megawatts are achieved via paralleling of individual units.

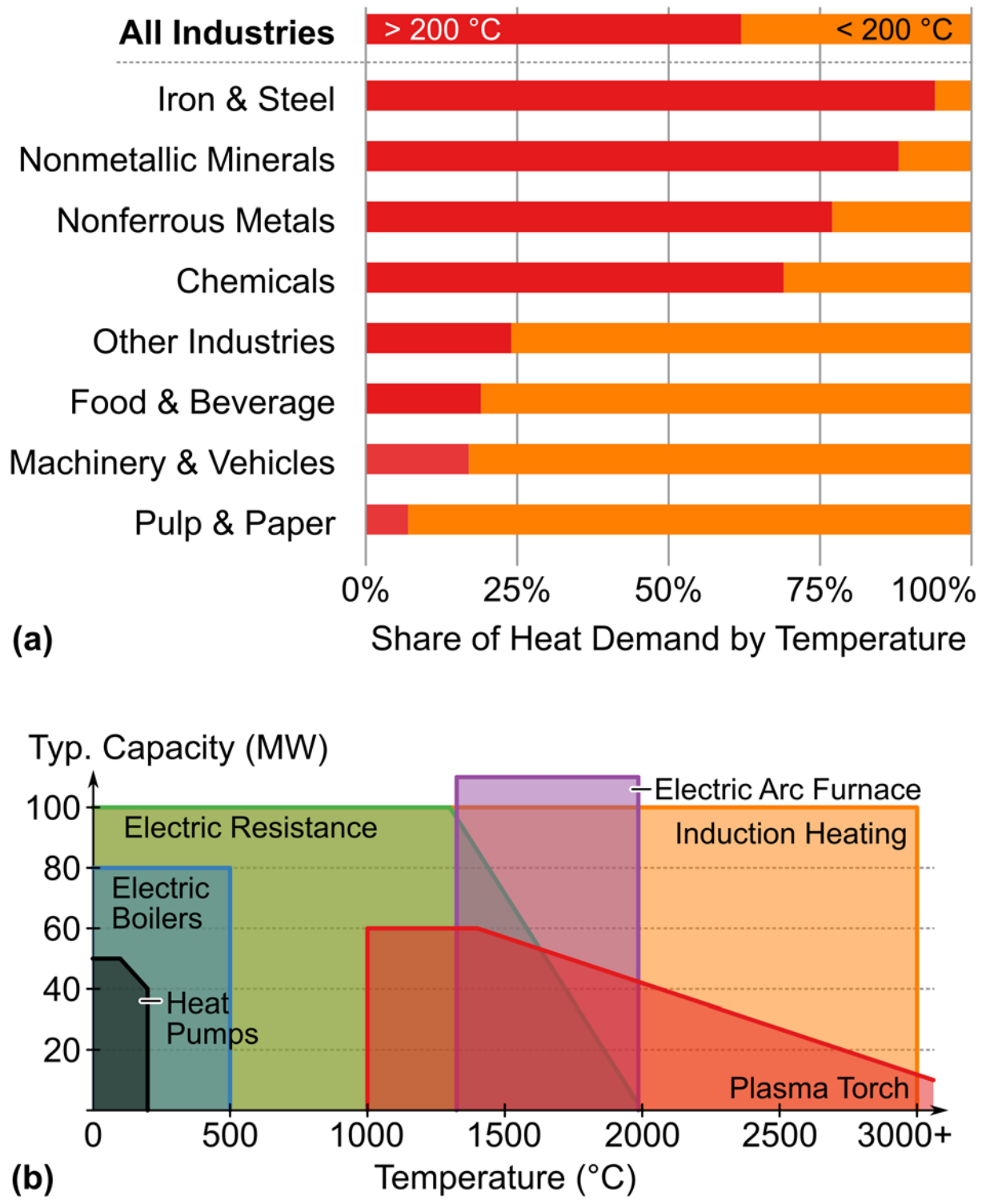

**Fig. 6. (a)** Share of heat demand at temperatures above 200 °C by industry sectors (data from [46]). **(b)** Electrified heating technologies with typical temperature ranges and typical power capacities expected by 2035 (data from [47]).

**Resistive Heating.** The first option to achieve temperatures beyond 200 °C at significant power levels is electric resistance heating based on the Joule effect, see **Fig. 6a**, providing rapid heating and accurate control of temperatures up to well beyond 1500 °C.

**Induction Heating.** Induction heating has a wide range of industrial applications and achieves even higher temperatures of up to 3000 °C, e.g., for metallurgy. Early patents date back to 1887 and, meanwhile, modern induction furnaces reach power levels of 20 MW, typically operating at the grid frequency of 50 Hz or 60 Hz [48]. The availability of high-power converters allows operating at higher frequencies of around 1 kHz, resulting in higher specific power of the furnace, and even higher operating frequencies are used for smaller furnaces, e.g., for precious metal processing [48].

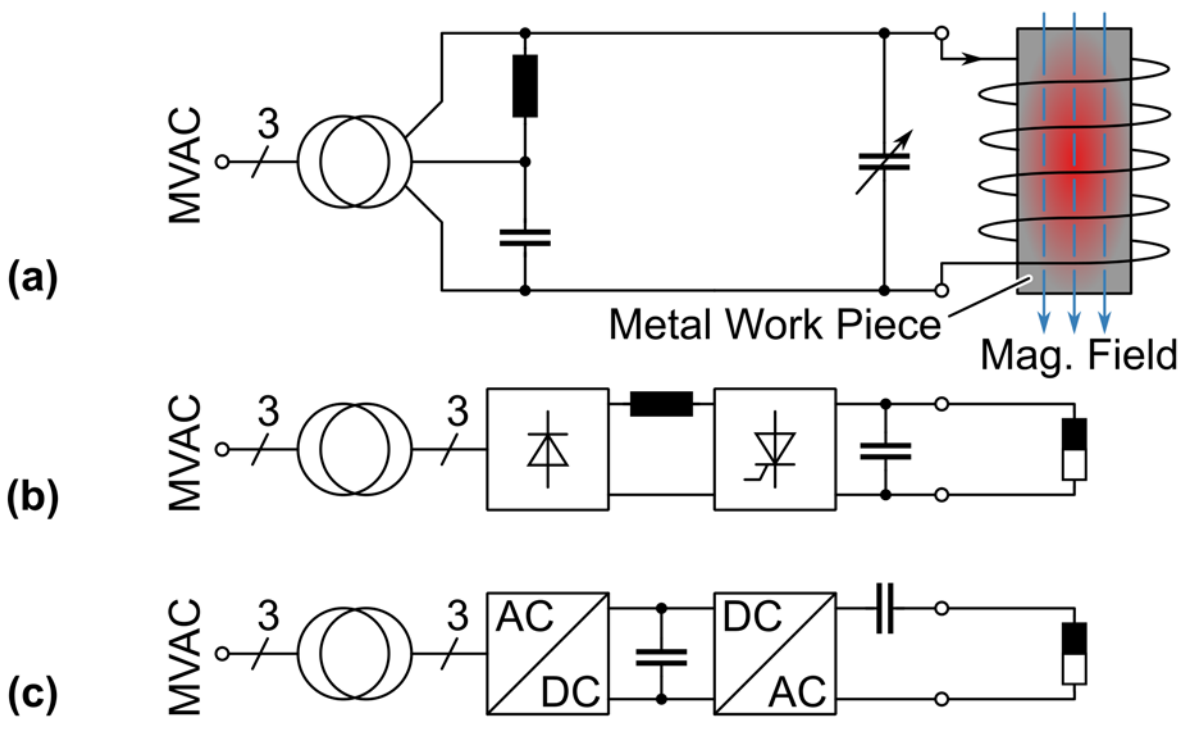

**Fig. 7.** Examples of induction heating power supplies. **(a)** Mains-frequency supply for high-power induction heating, using a Steinmetz-type symmetrizing circuit to interface the single-phase load to a three-phase grid and an adjustable capacitor for reactive power compensation of the inductive load. **(b)** Thyristor-based supply with a current dc-link and parallel compensation of the inductive load, and **(c)** voltage-source converter (realized with IGBTs or wide-bandgap power transistors) and series compensation of the inductive load. Figures based on [48].

The heating mechanism is based on eddy currents induced *inside* the material to be heated by an externally generated ac magnetic field, i.e., volumetric heating is achieved (in contrast to, e.g., resistive heating, where conductive or radiative heat transfer from the heating element to the material is needed). Therefore, the skin depth of the material influences the selection of the frequency and/or the sizing of the furnace [48]. In induction furnaces, electromagnetic forces act on molten metal and, advantageously cause a stirring effect [48]. There are also specialized induction heating application such as pulse induction hardening [49]. **Fig. 7** shows exemplary power supply topologies for high-power induction heating systems.

Induction heating of aluminum or copper billets (e.g., for forging or extrusion processes) with high conductivity is challenging, and efficiencies of induction heating with ac magnetic fields are limited to around 60% [49]. Thus, an alternative induction heating concept (termed “DC induction heating”) based on rotating the aluminum billets in a very high dc magnetic field generated by a coil made of high-temperature superconducting (HTS) tape was proposed around 2005 [50]. Recently, the technology has been scaled up, and a 1 MW demonstration system reached an efficiency of more than 80% [51].

Finally, heating nonconductive materials or fluids requires conductive heating elements placed in the fluid inside of a nonconductive vessel; recent research has investigated, e.g., thermochemical reactors using meta-material baffles and megahertz frequencies for uniform heating profiles [52].

**Plasma Torches.** Plasma torches heat a process gas to extreme temperatures up to 5000 °C and more, using an electric arc between two electrodes [26], [47]. In addition to the possibility of reaching extreme temperatures, plasma torches feature high energy density and small installation volumes, high power density of the ejected hot gas jet, free selection of the process gas, and fast start/stop dynamics, making them viable replacements for fossil fuel burners. Challenges are related to cooling and electrode lifetime due to the extreme conditions [47]. Applications of thermal plasmas (metallurgy, spray coating, chemicals, etc.) as well as generation methods (dc, ac, radio frequency, microwave, etc.) are far more than what can be covered here. Interested readers are referred to [26], which also covers electric arc furnaces discussed next. See **Fig. 4** and the "Steel" section above for a brief discussion on power supply requirements. Finally, note that there is ongoing research on *ac* plasma torches with possible advantages regarding better efficiency and lifetime [47].

**Electric Arc Furnaces.** Whereas the heat transfer of a plasma torch occurs via a carrier gas (the electric arc burns inside of the torch only), an EAF uses so-called transferred arcs, i.e., the arc burns between an electrode and the material to be heated itself (see also **Fig. 2c** and discussion there). This improves the heat transfer to the material by about an order of magnitude compared to a plasma torch, and the overall efficiency can exceed 90%, especially for melting processes where the high arc current flowing through the melt contributes with resistive heating [26], [47]. See **Fig. 3** and the "Steel" section for a brief discussion on power supply requirements.

**Dielectric Heating.** Microwave heating is employed in chemical and material processing, and in microwave-assisted chemistry [53]. The resulting direct, selective/local volumetric heating has been found to greatly increase reaction rates and yields, improve reaction completeness and selectivity, and enable solvent-free processes, resulting in lower energy consumption [53]. Whereas many applications of microwave heating use moderate temperatures, up to 2200 °C are achieved in the heat treating and sintering of ceramics [47].

Dielectric heating is based on dipole molecules in a material rotating in an applied radio-frequency electromagnetic field [53]. Microwaves are electromagnetic waves in the frequency range between 300 MHz and 300 GHz (typical frequencies are 915 MHz and 2450 MHz). Microwaves are generated using magnetrons, which are available with a wide range of power levels from 100s of watts to several megawatts (mainly for military radar applications), and show relatively high conversion efficiencies of up to 85% but suffer from limited lifetimes. **Fig. 8** shows the three power supplies required by an industrial magnetron. Klystrons reach even higher output power levels and much longer lifetimes but are usually too expensive for industrial applications. More recently, the advent of GaN power semiconductor technology has boosted solid-state microwave generators that currently reach power levels of several kilowatts and show significantly longer service lives than magnetrons (10 years vs. 10,000 hours); further, the microwave frequency can be varied and adjusted precisely. Interested readers can find a much more detailed discussion in Chapter 6 of [53].

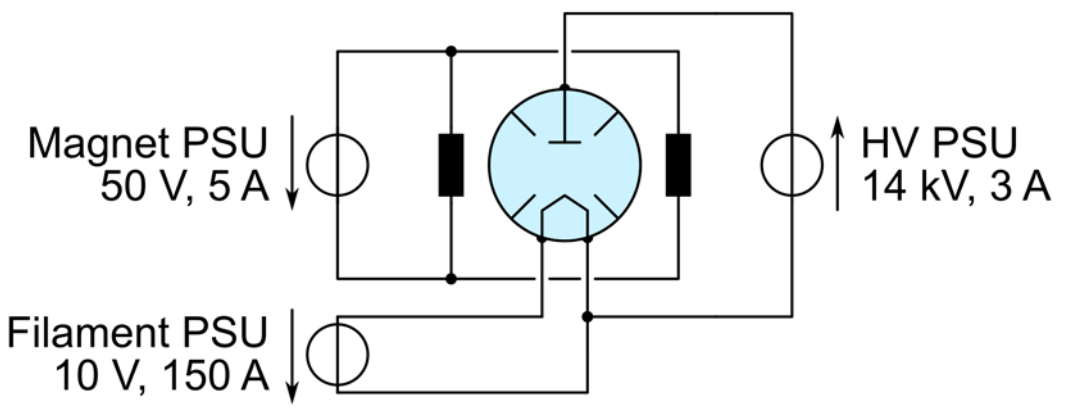

**Fig. 8.** Power supply requirements of an industrial magnetron in the 30 kW power range (figure adapted from [54]).

**Other Technologies.** There are other heating technologies for special applications, e.g., infrared heating, laser heating (e.g., for laser cutting), shock-wave heating, or heating by electron beams [47].

## E. Chemistry

Many chemical processes require (high-temperature) heat and benefit from volumetric heating (e.g., via induction or dielectric heating), which enables selective heating with high spatial and/or temporal resolution, enhancing chemical reactions via non-equilibrium thermal pathways with rapid heating/quenching sequences [55]. In the following, a few examples of other (not heating-related) electrification pathways researched in chemistry are highlighted.

**Plasma-Assisted Chemistry.** Non-thermal plasmas (the electron temperature is much hotter than the temperature of the heavier ions; in contrast to the thermal plasmas discussed above for heating applications) are used to enhance chemical reactions, e.g., for upcycling of the $CO_2$ in the flue gas of rotary kilns used for clinker production into a combustible mix of CO and $O_2$ [56], methane emission reduction in the exhaust from natural gas and dual-fuel engines in, e.g., marine applications [57], and in recycling processes [58]. Possible further applications include plasma-mediated water splitting for the production of hydrogen [59] and, more generally, plasma-assisted power-to-X processes [60]. As an aside, note that because only the electron temperature is high, non-thermal plasmas typically interact with gases or with surfaces of liquids and solids only, instead of with the bulk material. This, and the low (down to ambient) gas temperatures, makes them also applicable for medical purposes like inactivation of microorganisms, stimulation of angiogenesis, and cancer treatments [61], as well as many other applications like food preservation, etc. [62].

In general, plasma generation is an unstable process (positive feedback of temperature to current consumption with very fast dynamics), and hence very challenging requirements for the power converters such as very high regulation bandwidth and controlled output impedances arise [63], similar to those discussed above in the context of plasma torches.

**Electrochemical Processes.** Electrochemistry, i.e., the closing of an externally provided electric circuit through ion transfer between two electrodes in a solution, is a highly promising approach for the direct electrification of various emission-intensive chemical processes [65] such as steel production (see molten oxide electrolysis discussed above) or nitrogen fixation (ammonia production for fertilizer or as a synthetic fuel) with 50% lower $CO_2$ emissions and 75% lower energy usage [66], and also for more specialized applications like production of high-purity silicon [67], or recycling of electronic waste [68].

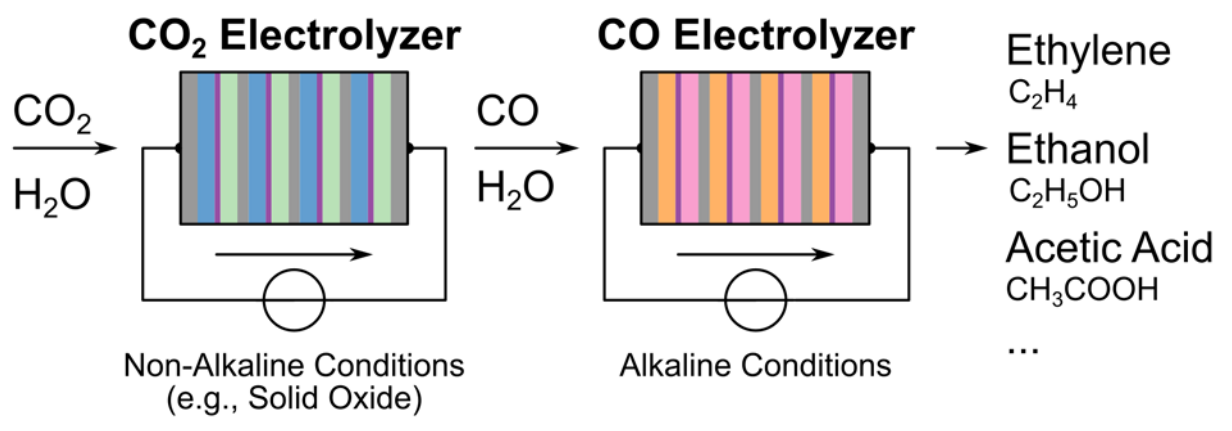

**Fig. 9.** Electrochemical conversion of $CO_2$ into multi-carbon feedstock molecules for the chemical industry (figure based on [64]).

As a further example, electrochemical reduction can be used to convert $CO_2$ to CO and then into multi-carbon molecules used as feedstock for a multitude of end products (see **Fig. 9**) [64], [65], [69], possibly using pulsed electrolysis [70]. Here, pulse durations down into the milliseconds timescale are used, and advantages of more complex voltage waveforms (compared to simple on/off schemes) are expected [71], implying specific requirements for high-current/low-voltage power supplies and/or benefits of co-designing these power supplies with the reactors. Similarly, pulsed proton exchange membrane water electrolysis can improve hydrogen production efficiency [16]. Finally, a recently proposed method for the decomposition of PFAS (per- and polyfluoroalkyl substances, commonly referred to as "forever chemicals") is envisioned to replace currently used alkaline metals as electron sources with an externally supplied electric current [72]. This illustrates how emerging ideas in chemistry might result in new applications for power converters with specific requirements.

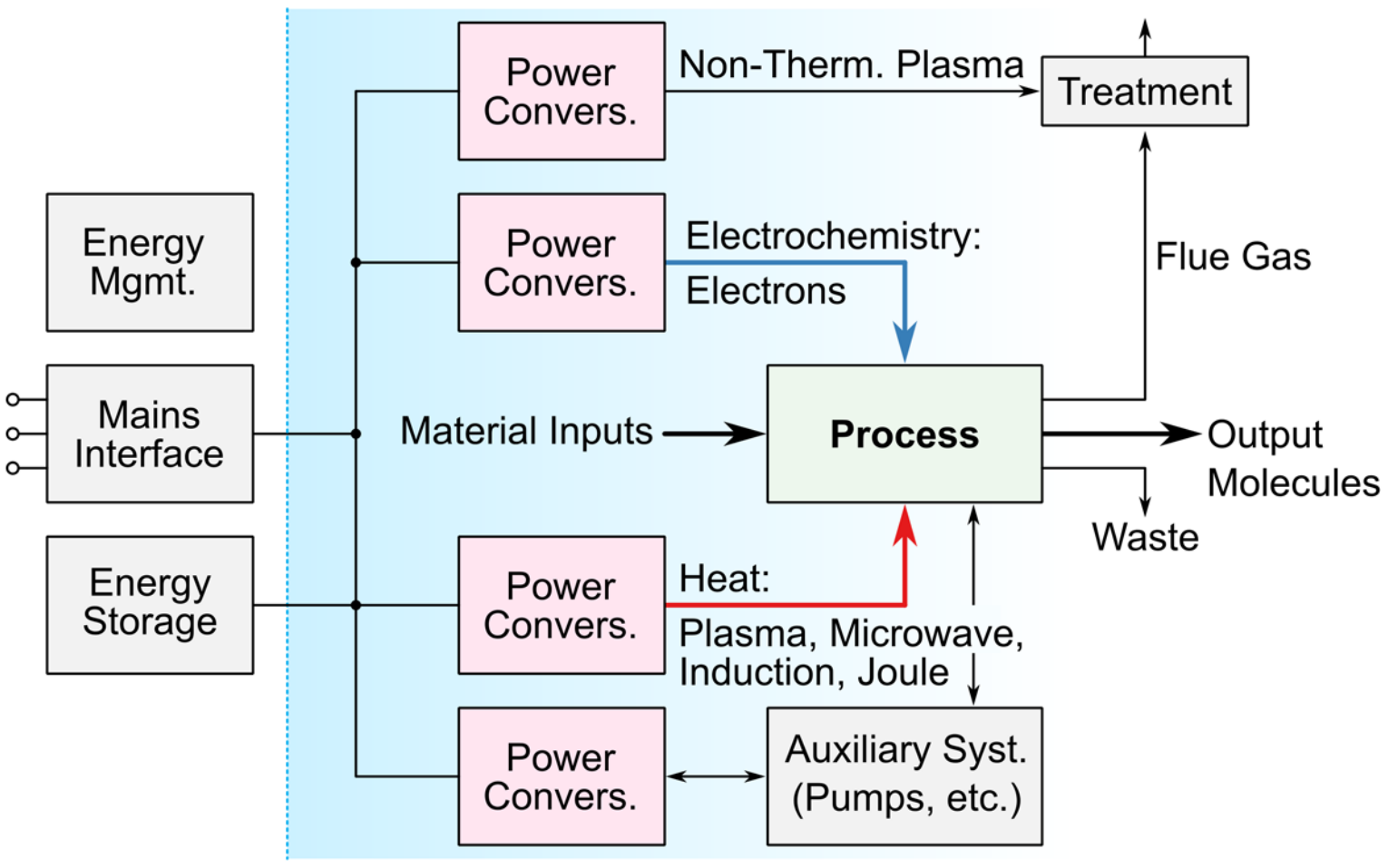

**Fig. 10.** Abstract representation of an industrial/chemical process. Power electronic conversion systems are expected to be instrumental in direct electrification of such processes.

## F. Summary

**Fig. 10** shows an abstract overview of an industrial process, typically chemical or metallurgical in nature, and illustrates the opportunities for future power electronic energy conversion systems to substantially enable the direct electrification of such processes, e.g., through electrochemical processes or the application of plasmas as discussed above. This results in a fascinating interdisciplinary research area at the interface between power electronics and process-specific science. Further research vectors should address system-level aspects such as grid-integration, mains interfaces, in-plant power distribution, and energy management (buffering of fluctuating power availability and demand). Finally, electrification pathways of auxiliary systems such as motor-drive pumps, etc. are relatively clear, yet still highly important.

# III. Conclusion and Outlook

Electrification of the hard-to-abate sectors is mandatory for reaching net zero carbon emission targets. Whereas indirect electrification via synthetic fuels will play a part (e.g., for shipping or aviation, and also for long-term energy storage in a future net-zero multi-carrier energy system [8]), *direct* electrification is preferable whenever possible. Therefore, the aim of this article was to give a broad overview of emerging technology options and research in neighboring disciplines such as chemistry, metallurgy, process engineering, etc.

Power electronics is a key technology for the electrification of *anything:* it is certain that power electronics will be essential for the emerging fields discussed above and others, but the specific requirements and research needs must still be fully identified. For example, plasma generation requires high-voltage power supplies with very fast control dynamics. Electrochemistry, on the other hand, requires very low output voltages and high currents, and pulsed operation poses questions regarding thermal management and reliability. Scaling up to industrially relevant dimensions requires strategies for processing very high power levels, and, finally, grid integration aspects and energy management must be studied.

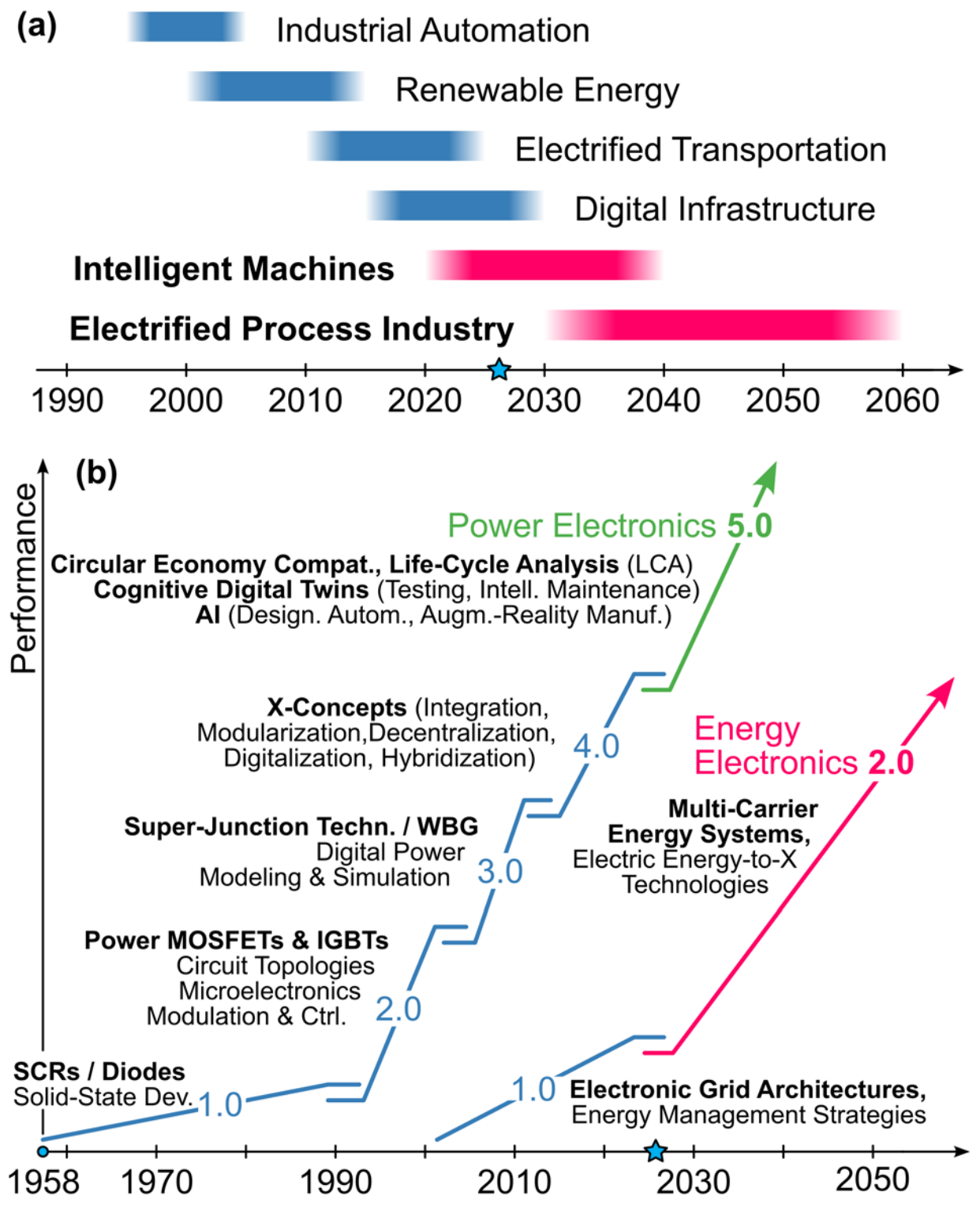

**Fig. 11. (a)** "Beyond-tomorrow" research challenges in power electronics over time: An exploratory phase initiates the ramp-up of research activities in new application areas; activities decrease again after handover to industry. **(b)** Technology S-Curves of power electronics and "energy electronics" with a broader scope including cross-sector energy conversion aspects [10].

Cutting-edge research at universities and also in industry should typically focus on "beyond-tomorrow" challenges, see **Fig. 11a**. Yet many of today's most active research areas are already well established in industrial practice. As we have argued in this article, we see the electrification of hard-to-abate sectors, in particular of the process industry, as the next upcoming highly relevant beyond-tomorrow research challenge in power electronics with a wide range of fascinating interdisciplinary research questions.

**Fig. 11b** shows the technology S-curves of power electronics. Historically driven by new device-level technologies like wide bandgap power semiconductors and new circuit topologies, etc., we recently argued that the next S-curve, termed "Power Electronics 5.0" will be defined by the need to take environmental compatibility and circular economy aspects into account [73]. This seems especially important given the broadening range of power electronics applications, for example for the electrification of the process industry. Similarly, a second set of S-curves could be defined to describe the development of *energy* electronics, i.e., with a broader scope that includes cross-sector energy conversion aspects like energy storage and, in the second wave, direct electrification of hard-to-abate sectors and/or power-to-X as discussed in this article.

Triggered by a 2025 white paper [3], a workshop on "Power Electronics in Future Process Industries: Electrification for Decarbonization" organized by the European Center for Power Electronics (ECPE) in fall 2026 aimed at connecting experts from different process industries with power electronics researchers to start addressing the “known unknowns” regarding the requirements for power converters supporting the vast array of possible electrification pathways: As power electronics researchers, we should go out and talk to the scientists and experts in the neighboring disciplines. Interdisciplinary research is required to clarify how optimum performance of future electrified industrial processes can be achieved, e.g., by co-developing and/or close integration of the power converter with the process-specific technology.

## About the Authors

***Jonas Huber*** (jh@ethz.ch) (Senior Member, IEEE) is a tenure-track assistant professor at ETH Zurich and leads the Power Electronics and Drive Systems Laboratory. His research interests comprise ultra-compact, ultra-efficient or highly dynamic power electronic and mechatronic systems and their environmental impacts, as well as emerging power electronics applications.

***Johann W. Kolar*** (kolar@ams.ee.ethz.ch) (Life Fellow, IEEE) is an International Member of the U.S. National Academy of Engineering and was a Full Professor and the Head of the Power Electronic Systems Laboratory at the Swiss Federal Institute of Technology (ETH) Zurich. As a Professor Emeritus (since August 2024), he continues to actively pursue research on ultra-compact and efficient WBG converter systems, AI applications in power electronics, Solid-State Transformers, and the life-cycle analyses of power electronics converter systems.